\documentclass[aps,prc,reprint,groupedaddress,showpacs]{revtex4-1}

\usepackage{bm}
\usepackage{longtable}
\usepackage{graphicx}

\begin{document}


\title{Fully coupled-channel complex scaling method for the $\bm{K^-pp}$ system}


\author{Akinobu Dot\'e}
\email[]{dote@post.kek.jp}
\altaffiliation{J-PARC Branch, KEK Theory Center, IPNS, KEK, 203-1, Shirakata, Tokai, Ibaraki, 319-1106, Japan}
\affiliation{KEK Theory Center, Institute of Particle and Nuclear Studies (IPNS), High Energy Accelerator Research Organization (KEK), 1-1 Oho, Tsukuba, Ibaraki, 305-0801, Japan}

\author{Takashi Inoue}
\affiliation{Nihon University, College of Bioresource Sciences, Fujisawa 252-0880, Japan}

\author{Takayuki Myo}
\affiliation{General Education, Faculty of Engineering, Osaka Institute of Technology, Osaka 535-8585, Japan}


\date{\today}

\begin{abstract}
We have developed a fully coupled-channel complex scaling method (ccCSM) for the study of the most essential kaonic nucleus, $``K^-pp,"$ which is a resonant state of a $\bar{K}NN$-$\pi\Sigma N$-$\pi\Lambda N$ coupled-channel system based on a theoretical viewpoint. By employing the ccCSM and imposing the correct boundary condition of resonance, the coupled-channel problem is completely solved using a phenomenological energy-independent potential. As a result of the ccCSM calculation of $``K^-pp,"$ in which all three channels are treated explicitly, we have obtained three-body resonance as a Gamow state. The resonance pole indicates that the binding energy of $``K^-pp"$ and the half value of its mesonic decay width are 51 MeV and 16 MeV, respectively. In the analysis of the resonant wave function obtained using the ccCSM, we clarify the spatial configuration and channel compositions of $``K^-pp."$ Compared with past studies of single-channel calculations based on effective $\bar{K}N$ potentials, the current study provides a guideline for the determination of the $\bar{K}N$ energy to be used in effective potentials.  
\end{abstract}


\maketitle


{\it Introduction:} Kaonic nuclei (nuclear systems with antikaons) are one of the important topics in hadron and strange nuclear physics because they exhibit several interesting properties that have never been observed in ordinary nuclei. The $\bar{K}N$ potential is considerably attractive, particularly in the isospin $I=0$ channel. It forms a quasi-bound state that corresponds to an excited hyperon, $\Lambda(1405)$ \cite{ChU:Review}. Early studies on phenomenological $\bar{K}N$ potential showed that light kaonic nuclei shrink significantly to form dense states, and a few of them exhibit interesting structures because of strong $\bar{K}N$ attraction \cite{AY_2002, AMDK}. Therefore, kaonic nuclei are expected to be a doorway to dense matter, in which the partial restoration of chiral symmetry may occur \cite{ChSymRes:Hatsuda, ChSymRes:Weise}. 

To describe the nature of kaonic nuclei in detail, considerable effort has been made toward the study of the most essential kaonic nucleus, $K^-pp$, in theoretical and experimental aspects. In the theoretical aspect, as $K^-pp$ is a three-body system, it has been investigated using various approaches. Typically,  the variational approach or Faddeev-AGS approach have been applied using phenomenological $\bar{K}N$ potentials or chiral SU(3)-based $\bar{K}N$ potentials.  As stated in Ref. \cite{StrangenessSummary_2016}, all theoretical studies show that $K^-pp$ can be bound and its binding energy should be less than 100 MeV. However, the binding energy of $K^-pp$ depends strongly on the type of $\bar{K}N$ potential used in calculation: When energy-independent phenomenological $\bar{K}N$ potentials are employed, $K^-pp$ is a relatively deeply-bound state \cite{Kpp:AY, Faddeev:Shevchenko}. On the contrary, it is a shallowly-bound state when energy-dependent chiral SU(3)-based potentials are employed \cite{Kpp:DHW, Kpp:IKS, Kpp:BGL}. In the experimental aspect, interesting results have been reported through the observation of a few signals, even though the signals have not yet been established as the $K^-pp$ bound state \cite{Kpp:exp_DISTO, Kpp-ex:JPARC-E27, Kpp-ex:JPARC-E15}. It should be noted that J-PARC E15 group is now finalizing the analysis of the data acquired in their second run: an exclusive measurement of the $^3$He($K^-$, $\Lambda p$)$n_{missing}$ reaction with high statistics \cite{Kpp-ex:JPARC-E15-2nd}. It is expected that the new result will provide a conclusive answer for the existence of the $K^-pp$ bound state through experimental observation.  

We have investigated $K^-pp$ using a coupled-channel complex scaling method (ccCSM). According to the abovementioned theoretical studies, $K^-pp$ should exist as a resonant state, and not as purely a bound state, between the $\pi\Sigma N$ and $\bar{K}NN$ thresholds because its binding energy is less than 100 MeV. In addition, $K^-pp$ is expected to be a coupled-channel system of the $\bar{K}NN$, $\pi\Sigma N$, and $\pi\Lambda N$ channels. This is similar to $\Lambda(1405)$, which is reasonably understood as a $\bar{K}N$-$\pi\Sigma$ coupled-channel system. Therefore, we consider that the treatments of the resonance and coupled-channel problems are the key factors in the theoretical study of $K^-pp$. The ccCSM can handle both factors simultaneously because it is based on the complex scaling method (CSM), which has been applied successfully in multiple studies of the resonances of ordinary nuclei, particularly those of unstable nuclei \cite{CSM:Myo2}. 

In a previous study, we have studied $K^-pp$ using a method based on the ccCSM, referred to as the ccCSM+Feshbach method \cite{ccCSM+F:Dote}. In this method, the coupled-channel problem of $K^-pp$ is reduced to a single-channel problem of $\bar{K}NN$ through the Feshbach projection, which is well realized on the CSM. Therefore, the advantage of the ccCSM+Feshbach method is that it reduces the computational cost. However, the dynamics of eliminated channels are lost in calculation, and it is impossible to obtain information about these channels from solutions. In other words, the nature of $K^-pp$ cannot be determined completely. To develop a complete understanding of $K^-pp$, we have attempted to use the {\it fully} ccCSM calculation, in which all channels are treated explicitly. In this method, we can directly obtain a resonant wave function so that the information about every channel is investigated explicitly.


{\it Formalism:} In this article, similarly to most earlier works, we consider the $K^-pp$ system as a $\bar{K}NN$-$\pi\Sigma N$-$\pi\Lambda N$ coupled-channel system with the following quantum numbers: total spin parity $J^\pi=0^-$, total isospin $T=1/2$, and $T_z=1/2$. Hereafter, such a $K^-pp$ state is symbolically denoted as $``K^-pp"$ ($K^-pp$ with double quotation marks). For the fully coupled-channel calculation, the $``K^-pp"$ wave function is expanded in terms of a basis as follows: 
\begin{widetext}
\begin{equation}
|``K^-pp"\rangle 
\; = \; \sum_{ch=1}^8 \sum_{n=1}^N \; C_n^{(ch)} \, F_n^{(ch)}(\bm{x}_1, \bm{x}_2) \; \left|S_{B_1 B_2 (ch)}=0 \right\rangle \; \left| T=1/2, T_z=1/2; \, \textrm{Isospin-Flavor}_{(ch)} \right\rangle. \label{K-pp_wfnc}
\end{equation}
\end{widetext}
Basically, the structure of the wave function is the same as that used in the ccCSM+Feshbach calculation, in which only the $\bar{K}NN$ channel was directly considered. (See Eqs.~(11) and (12) in Ref. \cite{ccCSM+F:Dote}.) In the current study, the $\pi\Sigma N$, $\pi\Lambda N$ and the $\bar{K}NN$ channels are treated explicitly. The 8 channels that can be coupled with the $``K^-pp"$ are listed in Table \ref{K-pp_wfnc_detail}. Channels $ch=1$ and 2 are the $\bar{K}NN$ channels, which are the same as those used in our previous study. Newly added channels, i.e., $ch=3,4, 5, 6$ and $ch=7, 8$ are the $\pi\Sigma N$ and $\pi\Lambda N$ channels, respectively. 

The coefficients, $\{C_n^{(ch)}\}$, are parameters to be determined by the diagonalization of the complex-scaled Hamiltonian matrix. Suffix $n$ is for the basis functions that expand the spatial part of the $``K^-pp"$ wave function, and it increases up to $N$ in each channel. The spatial basis function, $F_n^{(ch)}(\bm{x}_1, \bm{x}_2)$, is composed of a correlated Gaussian function, $G_n^{(ch)}(\bm{x}_1, \bm{x}_2)$. It is projected onto a parity eigenstate for the baryon-baryon system in each channel as $G_n^{(ch)}(\bm{x}_1, \bm{x}_2)\pm{G_n^{(ch)}}(-\bm{x}_1, \bm{x}_2)$. Here, Jacobi coordinates $\bm{x}_1$ and $\bm{x}_2$ are defined as $\bm{x}_1=\bm{r}_{B2}-\bm{r}_{B1}$ and $\bm{x}_2=\bm{r}_{M}-\bm{R}_{B1,B2}$. In other words, $\bm{x}_1$ and $\bm{x}_2$ correspond to the relative coordinate between two baryons ($B_1$ and $B_2$) and that between the meson ($M$) and center of mass of the baryons, respectively. For more details on the spatial part of the wave function, readers can refer to our previous paper \cite{ccCSM+F:Dote}. The total spin of the baryons, $S_{B_1 B_2 (ch)}$, is fixed at zero in all channels. The total isospin and its projection are assumed to be $T=1/2$ and $T_z=1/2$, respectively, and its structure is given in the last column of Table \ref{K-pp_wfnc_detail}. Here, it should be noted that in the $\pi\Sigma N$ and $\pi\Lambda N$ channels, the isospin part of the wave function must be symmetrized or antisymmetrized for baryon labels (referred to as {\it flavors}) to satisfy the generalized Pauli principle. 

\begin{table*}
\caption{Each channel of the $``K^-pp"$ wave function. 
``Spatial function $F_n^{(ch)}(\bm{x}_1, \bm{x}_2)$," 
``Spin $S_{B_1 B_2 (ch)}$," and ``Isospin-Flavor$_{(ch)}$" correspond to 
those given in Eq.~(\ref{K-pp_wfnc}). Function $G_n^{(ch)}(\bm{x}_1, \bm{x}_2)$ is a correlated Gaussian function. Symbols $\{YN\}_{Sym.}$ and $\{YN\}_{Asym.}$ ($Y=\Lambda, \Sigma$) in the last column indicate that the flavor labels of $YN$ are symmetrized and antisymmetrized, respectively. \label{K-pp_wfnc_detail}}
\begin{ruledtabular}
\begin{tabular}{l|ccc}
Channel ($ch$) & Spatial function $F_n^{(ch)}(\bm{x}_1, \bm{x}_2) \times \sqrt{2}$ & Spin $S_{B_1 B_2 (ch)}$ & Isospin-Flavor$_{(ch)}$\\
\hline
1 & $G_n^{(1)}(\bm{x}_1, \bm{x}_2)+{G_n^{(1)}}(-\bm{x}_1, \bm{x}_2)$ & $S_{NN}=0$ & $[\bar{K}[NN]_1]_{(1/2,1/2)}$ \\
2 & $G_n^{(2)}(\bm{x}_1, \bm{x}_2)-{G_n^{(2)}}(-\bm{x}_1, \bm{x}_2)$ & $S_{NN}=0$ & $[\bar{K}[NN]_0]_{(1/2,1/2)}$ \\
\hline
3 & $G_n^{(3)}(\bm{x}_1, \bm{x}_2)+{G_n^{(3)}}(-\bm{x}_1, \bm{x}_2)$ & $S_{\Sigma N}=0$ & $[[\pi\Sigma]_0 N]_{(1/2,1/2)}$, $\{\Sigma N\}_{Sym.}$ \\
4 & $G_n^{(4)}(\bm{x}_1, \bm{x}_2)-{G_n^{(4)}}(-\bm{x}_1, \bm{x}_2)$ & $S_{\Sigma N}=0$ & $[[\pi\Sigma]_0 N]_{(1/2,1/2)}$, $\{\Sigma N\}_{Asym.}$ \\
5 & $G_n^{(5)}(\bm{x}_1, \bm{x}_2)+{G_n^{(5)}}(-\bm{x}_1, \bm{x}_2)$ & $S_{\Sigma N}=0$ & $[[\pi\Sigma]_1 N]_{(1/2,1/2)}$, $\{\Sigma N\}_{Sym.}$ \\
6 & $G_n^{(6)}(\bm{x}_1, \bm{x}_2)-{G_n^{(6)}}(-\bm{x}_1, \bm{x}_2)$ & $S_{\Sigma N}=0$ & $[[\pi\Sigma]_1 N]_{(1/2,1/2)}$, $\{\Sigma N\}_{Asym.}$ \\
\hline
7 & $G_n^{(7)}(\bm{x}_1, \bm{x}_2)+{G_n^{(7)}}(-\bm{x}_1, \bm{x}_2)$ & $S_{\Lambda N}=0$ & $[[\pi\Lambda]_1 N]_{(1/2,1/2)}$, $\{\Lambda N\}_{Sym.}$ \\
8 & $G_n^{(8)}(\bm{x}_1, \bm{x}_2)-{G_n^{(8)}}(-\bm{x}_1, \bm{x}_2)$ & $S_{\Lambda N}=0$ & $[[\pi\Lambda]_1 N]_{(1/2,1/2)}$, $\{\Lambda N\}_{Asym.}$ \\
\end{tabular}
\end{ruledtabular}
\end{table*}

Hamiltonian $\hat{H}$ is composed of a mass term, $\hat{M}$, a kinetic energy term, $\hat{T}$,  nucleon-nucleon ($NN$) potential, $\hat{V}_{NN}$, and a meson-baryon potential, $\hat{V}^{MB}_{\alpha \beta}$, for channels $\alpha$ and $\beta$, as follows: 
\begin{equation}
\hat{H} = \hat{M}+\hat{T} + \hat{V}_{NN} + \sum_{i=1,2} \sum_{\alpha,\beta=\bar{K}N, \pi\Sigma, \pi\Lambda} \hat{V}^{MB}_{\alpha \beta} (M,B_i). \label{Hamiltonian}
\end{equation}
The kinetic energy term, $\hat{T}$, is constructed for the Jacobi coordinates $(\bm{x}_1, \bm{x}_2)$ in each channel, which is similar to our previous study \cite{ccCSM+F:Dote}. The Argonne v18 potential \cite{Av18} is employed as the $NN$ potential. A phenomenological $\bar{K}N$ potential (referred to as the AY potential \cite{AY_2002}) is employed as the $\bar{K}N$ potential in the meson-baryon potential. The AY potential is constructed for a $\bar{K}N$-$\pi Y$ coupled-channel space, and it is given in $r$-space local form with a Gaussian shape. Its parameters are constrained with low-energy $\bar{K}N$ scattering data and the pole position of $\Lambda(1405)$. (Details are explained in Ref. \cite{AY_2002}.)  Note that the channel coupling of the AY potential is explicitly included in the Hamiltonian in Eq.~(\ref{Hamiltonian}).
In the present study, the $YN$ and $\pi N$ potentials are neglected because their contribution to the ``$K^-pp$'' energy is considered to be minor compared to that of the $NN$ and $\bar{K}N$ potentials. 

We apply the CSM to obtain resonances directly using the wave function defined in Eq.~(\ref{K-pp_wfnc}) \cite{CSM:Myo2}. In the CSM, all coordinates included in Hamiltonian $\hat{H}$ are complex-scaled as $\bm{x}_i \rightarrow \bm{x}_i e^{i \theta}$ with a common scaling angle, $\theta$. By diagonalizing complex-scaled Hamiltonian $\hat{H}^\theta$ using the basis functions given in Eq.~(\ref{K-pp_wfnc}), all eigenstates are obtained in discretized form. Among these states, the resonance states of $``K^-pp"$ are associated with complex-energy eigenvalues, which are independent of scaling angle $\theta$. For such complex-energy eigenvalues, each vector of complex coefficients $\{C_n^{(ch)}\}$ represents the corresponding resonance state.

We comment on the symmetry for the exchange of two baryons in $``K^-pp."$ As shown in the last term of Eq.~(\ref{Hamiltonian}), the meson-baryon potential is common for baryons $B_1$ and $B_2$ in all channels. In other words, the Hamiltonian is symmetric under the exchange of two baryons. On the other hand, the $``K^-pp"$ wave function given as Eq.~(\ref{K-pp_wfnc}) is antisymmetric for the baryon exchange in every channel, as explained above. Therefore, as mentioned in an early study on Faddeev calculation involving different kinds of baryons \cite{Faddeev:Miyagawa}, the wave function is antisymmetric for baryon exchange, whereas the Hamiltonian is symmetric.


\begin{figure}
\includegraphics[width=0.47\textwidth]{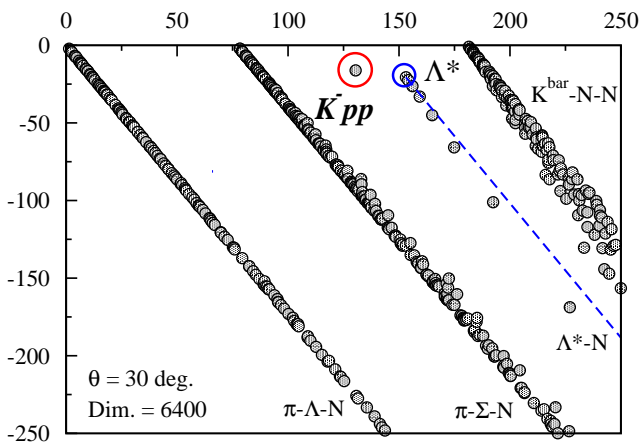}%
\caption{(Color online) Distribution of the complex-energy eigenvalues obtained for the $``K^-pp"$ system. The horizontal axis shows the real energy measured from the $\pi\Lambda N$ threshold, and the vertical axis shows imaginary energy. All values of energy are given in MeV. The eigenvalues marked with red and blue colored circles correspond to the $``K^-pp"$ and $\Lambda(1405)$ poles, respectively. Note that the thresholds of $\pi\Lambda N$, $\pi\Sigma N$, and $\bar{K}NN$ are located at 0 MeV, 77 MeV, and 181 MeV on the horizontal axis, respectively. \label{Eigenvalues_ccCSM}}
\end{figure}

{\it Result and discussion:} Fig.~\ref{Eigenvalues_ccCSM} shows distribution of the eigenvalues on the complex energy plane, obtained in the present ccCSM calculation by diagonalizing the complex-scaled Hamiltonian matrix. Here, scaling angle $\theta$ is set to be 30 degrees. We consider 20 Gaussian basis functions for individual Jacobi coordinates in each channel, and the total number of basis functions is 6400, in which all sets of Jacobi coordinates are considered. In the figure, the origin corresponds to the $\pi\Lambda N$ threshold. Hereafter, to represent the complex energy of poles, we use ``$-B_{K^-pp}$,'' which denotes the real energy measured from the $\bar{K}NN$ threshold, and ``$-\Gamma_{\pi YN}/2$,'' which denotes the imaginary energy and corresponds to the half value of mesonic decay width with a minus sign. In the CSM, the eigenvalues of continuum states appear along a $2\theta$ line, whose slope angle ($-2\theta$) depends on scaling angle $\theta$ \cite{CSM:Myo2}. In Fig.~\ref{Eigenvalues_ccCSM}, we clearly obtain the $\pi\Lambda N$, $\pi\Sigma N$, and $\bar{K}NN$ three-body continuum states along three $2\theta$ lines starting from the individual energy thresholds. In addition, there is a group of eigenvalues along a line starting from a complex eigenvalue of $(-B_{K^-pp}, -\Gamma_{\pi YN}/2)=(-28, -20)$ MeV (blue dashed line in Fig.~\ref{Eigenvalues_ccCSM}).  The complex energy at the starting point of this line (marked with a blue circle) coincides with the pole position of $\Lambda(1405)$ calculated using the AY potential \cite{AY_2002}. Therefore, this group of the eigenvalues represent the $\Lambda(1405)$-$N$ quasi two-body continuum state. Finally, we determine an eigenvalue that exists in addition to all lines mentioned above. This eigenvalue corresponds to the three-body resonance pole of the $``K^-pp"$ system. Based on this result, the binding energy and half decay width of the $``K^-pp"$ resonant state are determined to be 
\begin{equation}
B_{K^-pp}=51 \, {\rm MeV \quad and} \quad \Gamma_{\pi YN}/2=16 \, {\rm MeV}, 
\label{Pole_K-pp_ccCSM}
\end{equation}
respectively, when the Argonne v18 $NN$ potential and AY $\bar{K}N$ potential are employed. We have confirmed that the pole positions of the $``K^-pp"$ system and $\Lambda(1405)$ are considerably stable when scaling angle $\theta$ and the parameters of the spatial part of basis functions $\{ G_n^{(ch)}(\bm{x}_1, \bm{x}_2) \}$ change. It is noted that the eigenstates shown in Fig.~\ref{Eigenvalues_ccCSM} construct the completeness relation consisting of the three-body resonance and the two- and three-body continuum-states of each channel including $\Lambda(1405)$ \cite{CSM:Myo2}. This completeness relation is a unique characteristic of the ccCSM, and it is useful for applying the present wave function to spectrum calculation \cite{CSM:Myo2}. 

\begin{table}
\caption{Properties of the $K^-pp$ resonant state. The left column shows the norm of each component in $K^-pp$; ``${\cal N}(\bar{K}NN)$'' denotes the norm of the $\bar{K}NN$ component, and so on. The right column shows the mean distance between particles in $\bar{K}NN$ channel. ``$R_{NN}$'' and ``$R_{\bar{K}N(I=0,1)}$'' are the $NN$ and isospin-decomposed $\bar{K}N$ distances, respectively. \label{Prop_K-pp}}
\begin{ruledtabular}
\begin{tabular}{lrr|lr}
${\cal N}(\bar{K}NN)$ & $1.004-i0.286$    & & $R_{NN}$ & $1.86+i0.14$ [fm]\\
${\cal N}(\pi\Sigma N)$ & $-0.002+i0.276$ & & $R_{\bar{K}N(I=0)}$ & $1.40-i0.01$ [fm]\\
${\cal N}(\pi\Lambda N)$ & $-0.002+i0.010$ & & $R_{\bar{K}N(I=1)}$ & $1.95+i0.14$ [fm]
\end{tabular}
\end{ruledtabular}
\end{table}

Table \ref{Prop_K-pp} shows the properties of the $K^-pp$ resonant state, which are determined using the ccCSM wave function. The norm of each component is given in the left column. Although the quantities for resonant states are obtained to be complex values in the CSM, the magnitude of each norm indicates that the $\bar{K}NN$ component is apparently dominant and that the $\pi\Sigma N$ component is significantly involved in the $K^-pp$ resonance. This is attributed to the nature of the AY potential, in which the $\bar{K}N$-$\pi\Sigma$ coupling potential is considerably strong, particularly in the $I=0$ channel \cite{AY_2002}. A similar value of the $\pi\Sigma$ component is observed in $\Lambda(1405)$, which is an $I=0$ $\bar{K}N$-$\pi\Sigma$ coupled-channel system, for which ${\cal N}(\bar{K}N)=1.118-i0.107$ and ${\cal N}(\pi\Sigma)=-0.118+i0.107$. The mean $NN$ and $\bar{K}N$ distances in the $K^-pp$ system are shown in the right column. The $NN$ distance is calculated to be 1.86 fm. In nuclear matter with normal density ($\rho_0=0.16$ fm$^{-3}$), the mean distance between two nucleons is estimated to be 2.2 fm. This implies that the $NN$ distance in ``$K^-pp$'' decreases by 15\% compared with that in normal nuclear matter. Therefore, the $K^-pp$ system calculated using the AY potential could be a piece of dense matter \cite{Kpp:AY}. (The density is estimated to be $1.7\rho_0$ using the $NN$ distance.) The $\bar{K}N$ mean distance for the $I=0$ component is apparently smaller than that for the $I=1$ component because the $\bar{K}N$ potential is considerably more attractive in the $I=0$ channel than it is in the $I=1$ channel. In addition, the $I=0$ $\bar{K}N$ distance in the $K^-pp$ system is close to that in $\Lambda(1405)$. (The $\bar{K}N$ distance in $\Lambda(1405)$ is calculated to be $1.25 - i0.27$ fm in the case of the AY potential.) Therefore, the $\Lambda(1405)$ component is considered to survive in the $K^-pp$ system, as suggested in earlier studies \cite{ccCSM+F:Dote, Kpp:AY, Kpp:DHW}. 

\begin{table}
\caption{Comparison with past calculations of the $K^-pp$ system. In the column labeled ``Method,'' ``CSM'' represents ordinary CSM for a single-channel case and ``ccCSM+F'' represents the ccCSM+Feshbach method \cite{ccCSM+F:Dote}. In the column labeled ``AY pot.~type,'' ``Single'' and ``Coupled'' denote the single-channel \cite{Kpp:AY} and coupled-channel \cite{AY_2002} versions of the AY potential \cite{AY_2002}. ``Ansatz'' is used in the ccCSM+Feshbach method 
to estimate the $\bar{K}N$ energy in $\bar{K}NN$. (Refer to the text.) 
``$B_{K^-pp}$'' and ``$\Gamma_{\pi YN}/2$'' denote the binding energy and half decay width of the $K^-pp$ system, respectively. They are expressed in MeV. 
\label{Comp_Kpp}}
\begin{ruledtabular}
\begin{tabular}{lccc|cc}
Method & AY pot. type & Ansatz & & $B_{K^-pp}$ & $\Gamma_{\pi YN}/2$ \\
\hline
CSM      & Single & ---               & & 45 & 30 \\
\hline
ccCSM+F  & Coupled & $\Lambda^*$ fixed & & 45 & 29 \\
  &  & Particle   & & 46 & 27 \\
  &  & Field      & & 49 & 17 \\
\hline
ccCSM    & Coupled & ---  & & 51 & 16 \\
\end{tabular}
\end{ruledtabular}
\end{table}

The binding energy and decay width shown in Eq.~(\ref{Pole_K-pp_ccCSM}) indicate the genuine pole position of the $``K^-pp"$ resonance for the Hamiltonian (\ref{Hamiltonian}) with the AY potential because all channels are treated explicitly in the present ccCSM calculation. Therefore, it is interesting to compare the result of the present ccCSM calculation with those of earlier studies. In most previous studies \cite{Kpp:AY, Kpp:DHW, Kpp:BGL} and in our study on the ccCSM+Feshbach method, three-body calculation is carried out in the scheme of the $\bar{K}NN$ channel using only effective $\bar{K}N$ potential, in which the dynamics of the $\pi Y$ channels are incorporated indirectly. In these truncated calculations, the derived effective potential is energy dependent because of channel elimination, even though the original coupled-channel potential is energy independent. This energy dependence requires self-consistency for the $\bar{K}N$ energy while solving the Schr\"odinger equation. However, in principle, it is impossible to uniquely determine the $\bar{K}N$ energy in the $\bar{K}NN$ three-body system because subsystem energy is not an eigenvalue of the total Hamiltonian. To estimate the $\bar{K}N$ energy in the $\bar{K}NN$ system, two extreme ansatzes were proposed in a previous study, which were referred to as Field picture and Particle picture \cite{Kpp:DHW}. Using the binding energy, $B_K$, of an antikaon, the $\bar{K}N$ energy, $E_{KN}$, is estimated as $E_{KN}=-B_K$ in Field picture and $E_{KN}=-B_K/2$ in Particle picture.  (Details are explained in Ref. \cite{Kpp:DHW}.) According to the systematic study of light kaonic nuclei using the stochastic variational method with effective $\bar{K}N$ potentials \cite{SVM:S-Ohnishi}, the discrepancy among the results obtained by employing the two ansatzes increases with the mass of systems, particularly in the case of decay widths. Therefore, ambiguity exists in single-channel calculations performed using effective $\bar{K}N$ potential.  

Table \ref{Comp_Kpp} shows the binding energy and half decay width of the $``K^-pp"$ system, which are calculated using three methods based on the CSM with the AY potential (single/coupled channel version). The first row shows the result obtained using the CSM with the single-channel version of the AY potential, which is tuned to reproduce the $\Lambda(1405)$ energy calculated using the coupled-channel version of the AY potential. The results shown in the second, third, and fourth rows are obtained using the ccCSM+Feshbach method, in which three ansatzes are employed for the estimation of the $\bar{K}N$ energy. In addition to the two previously mentioned ansatzes, another ansatz (``$\Lambda^*$ fixed'') is examined, in which the $\bar{K}N$ energy is equal to that of $\Lambda(1405)$. The last row shows the result of the present ccCSM calculation. Among these calculations, the CSM with the single-channel potential and the ccCSM+Feshbach method with the $\Lambda^*$-fixed ansatz are conceptually equivalent. Actually, their results are in good agreement with each other. Furthermore, the result obtained using ccCSM+Feshbach method with the Particle-picture ansatz is similar to these two results. Based on this, the Particle-picture ansatz is considered to be almost equivalent to the $\Lambda^*$-fixed ansatz. On the contrary, the result obtained using the Field-picture ansatz is different from those mentioned above, particularly for decay width. However, it is considerably close to the result of the present ccCSM calculation, which is a fully coupled-channel calculation. Based on these comparisons, it can be concluded that Field picture is better than Particle picture when effective $\bar{K}N$ potential is used. Needless to say, it is best to carry out a fully coupled-channel calculation, such as that performed in this study.

Here, we comment further on the treatment of the $\bar{K}N$ energy. A more sophisticated ansatz has been proposed for it in an earlier study of $\bar{K}NN$ conducted using the hyperspherical harmonics approach \cite{Kpp:BGL}. This ansatz is considered to be an improved version of Particle picture. The $\bar{K}N$ energy is estimated as $E_{KN}=-B_K/2-\Delta$, in which $\Delta$ represents a correction. When this improved ansatz is employed in the ccCSM+Feshbach method,  the result is almost equal to that obtained using Field picture \cite{ccCSM+F:Dote(MENU2016)}. Therefore, even if we employ Particle picture, considering its improvement, the result becomes equal to that obtained using Field picture. This fact supports Field picture rather than Particle picture, consistently with the conclusion described in the previous paragraph.


{\it Summary and future plan:} We have developed a {\it fully} coupled-channel complex scaling method (ccCSM) for studying the most essential kaonic nucleus $``K^-pp"$. Theoretically, $``K^-pp"$ is regarded as a resonant state consisting of the $\bar{K}NN$-$\pi\Sigma N$-$\pi\Lambda N$ coupled-channel system. 
%
%
As the development of our previous study on the ccCSM+Feshbach method, the $\bar{K}NN$, $\pi\Sigma N$, and $\pi\Lambda N$ channels are treated explicitly in the present ccCSM calculation. As the first trial, we have employed a phenomenological $\bar{K}N$ potential (the AY potential). We have clearly obtained the $``K^-pp"$ three-body resonance pole. 
The resonance pole indicates that the binding energy is 51 MeV and the half width of the mesonic decay mode is 16 MeV. The analysis of the ccCSM wave function shows that the $``K^-pp"$ resonant state is apparently dominated by the $\bar{K}NN$ component and involves the $\pi\Sigma N$ component significantly. The mean distance between two nucleons is smaller than the value of normal density by approximately 15\%. Hence, we have confirmed that the $``K^-pp"$ system is a piece of a dense matter in the case of the AY potential, as suggested by earlier studies. As compared to previous studies, the current fully coupled-channel calculation provides a guideline on using effective $\bar{K}N$ potentials. For the determination of the $\bar{K}N$ energy in the total system, which is necessary in single-channel calculations using effective $\bar{K}N$ potentials, two ansatzes have been proposed in a previous study \cite{Kpp:DHW}. The present fully ccCSM calculation supports the Field-picture ansatz rather than the Particle-picture ansatz. 

Here, we emphasize that the present ccCSM calculation is the first calculation in which the ``$K^-pp$" three-body resonant wave function is obtained completely. This enables us to analyze the nature of the ``$K^-pp$" resonant state in detail. The $``K^-pp"$ resonance pole determined in the present calculation should be the genuine pole for the Hamiltonian (\ref{Hamiltonian}) involving  the AY potential. As the AY potential is energy independent, self-consistency for the $\bar{K}N$ energy is not necessary when all channels are treated explicitly. 
In the present case, by diagonalization of the complex-scaled Hamiltonian matrix, the resonance pole should definitely be obtained. 

In future, we will perform calculations using the fully ccCSM with chiral SU(3)-based $\bar{K}N$ potential. This is an energy-dependent potential, which is in contrast to the AY potential. Many studies based on chiral dynamics have suggested that $\Lambda(1405)$ exhibits a double-pole structure owing to the energy dependence of the potential \cite{ChU:Review}. In addition, in our previous study on the ccCSM+Feshbach method \cite{ccCSM+F:Dote(HYP2015)} and in an earlier study on the Faddeev-AGS approach \cite{Kpp:IKS}, it has been suggested that the $K^-pp$ system might exhibit a double-pole structure similar to $\Lambda(1405)$. As discussed in Ref. \cite{ccCSM+F:Dote(MENU2016)}, based on the ccCSM+Feshbach calculation, such double poles of the $K^-pp$ system are related to the experimental results mentioned at the beginning of this article: DISTO \cite{Kpp:exp_DISTO} and J-PARC E27 \cite{Kpp-ex:JPARC-E27} collaborations observed a signal close to the $\pi\Sigma N$ threshold, while J-PARC E15 \cite{Kpp-ex:JPARC-E15} collaboration observed a signal close to the $\bar{K}NN$ threshold. If these signals are regarded  as the $K^-pp$ bound state, the first two results indicate a deeply bound $K^-pp$ state, while the third indicates a shallowly bound $K^-pp$ state. One should recall the second run of the J-PARC E15 experiment \cite{Kpp-ex:JPARC-E15-2nd}. New data, which are currently being analyzed, would provide a definite conclusion about the $K^-pp$ bound state, as an exclusive measurement has been carried out with high statistics.
When all channels are treated explicitly in the fully ccCSM calculation, which is in contrast to our previous study \cite{ccCSM+F:Dote}, the difference between their nature can be determined, particularly the difference between their compositions. Such detailed information about $K^-pp$ will provide insights toward the understanding of the experimental results. Furthermore, two-nucleon absorption ($\bar{K}NN\rightarrow YN$), which is not accounted for in the current study, should be considered because it results in a large width of the non-mesonic decay mode \cite{2Nabs:Bayar-Oset}. Spectrum calculation is important for direct comparison with experimental results; this calculation is performed in Ref. \cite{JPARC-E15:Sekihara-Oset-Ramos}. In addition, such a study can be conducted using the fully ccCSM calculation with the completeness relation \cite{CSM:Myo2}. We expect that the fully ccCSM calculation with chiral SU(3)-based and AY $\bar{K}N$ potentials and experimental data will provide a conclusive answer to this longstanding issue of the most essential kaonic nucleus, $K^-pp$. 

{\it Acknowledgments:} One of the authors (A. D.) thanks Prof. T. Harada and Prof. H. Horiuchi for productive discussion on the treatment of the coupled-channel problem and Prof. Y. Akaishi for his helpful advice. This work is supported by JSPS KAKENHI Grant Number 25400286 and partially by Grant Numbers 24105008, 15K05091, and 26400281. The calculation in this study was performed using High Performance Computing system (miho) at Research Center for Nuclear Physics (RCNP) in Osaka University.  


\end{document}